\documentclass[12]{iopart}

\begin{document}

\title{Patterns of coexisting superconducting and particle-hole condensates}
\author{S. Tsonis$^1$, P. Kotetes$^1$, G. Varelogiannis$^2$ and P. B. Littlewood$^3$}
\address{$^{1,2}$ Department of Physics, National Technical University
of Athens, GR-15780 Athens, Greece}
\address{$^3$ Cavendish
Laboratory, University of Cambridge, Cambridge CB3 03E, United
Kingdom}

\ead{\mailto{$^1$pkotetes@central.ntua.gr}}
\ead{\mailto{$^2$varelogi@central.ntua.gr}}

\begin{abstract}
We have studied systematically the influence of particle-hole
symmetric and asymmetric kinetic terms on the ordered phases that we
may observe competing or coexisting in a tetragonal system. We show
that there are precise patterns of triplets of ordered phases that
are accessible (i.e. it is impossible to observe two of them without
the third one). We found a systematic way to predict these patterns
of states and tested it by identifying at least 16 different
patterns of three order parameters that necessarily coexist in the
presence of the kinetic terms. We show that there are two types of
general equations governing the competition of all these triplets of
order parameters and we provide them.
\end{abstract}

\newcommand{\be}[1]{\begin{equation}\label{eq:#1}}
\newcommand{\ee}{\end{equation}}
\newcommand{\bea}{\begin{eqnarray}}
\newcommand{\eea}{\end{eqnarray}}
\newcommand{\bm}{\mathbf}
\newcommand{\bt}{\textbf}
\newcommand{\mb}{\mathbf}
\newcommand{\phd}{\phantom{\dag}}
\newcommand{\ph}{\phantom{.}}
\newcommand{\noi}{\noindent}
\newcommand{\no}{\nonumber}

\submitto{\JPCM} \maketitle

\section{Introduction}

Almost all important functional materials undergo a pleiad of
phases that under certain conditions may coexist. Controlling the
parameters leading to coexistence is of primary importance as it
can provide access to new intriguing phenomena and
functionalities. Already in the early eighties, systematic
theoretical investigations of the coexistence of two ordered
electronic phases appeared
\cite{{Bibro},{Balseiro},{Machida1},MachidaMatsubara,{Machida2},
{Littlewood},{Grest},Birman,Psaltakis}, motivated essentially by
the general problem of antiferromagnetic superconductivity that
emerged in organic superconductors and heavy fermions. Numerous
theoretical studies of the coexistence and competition of two
phases continue to appear as the number of such experimental
paradigms multiplies.
\par It has been a general conclusion from all the above studies that
we must take into consideration {\it on the same footing the two
order parameters} (OPs) that compete and may eventually coexist,
otherwise we miss qualitatively new phenomena associated with this
competition. However, as we shall show below, additional order
parameters may coexist as well, and the need to include them is
equally important. In fact, we will show that the particle-hole
symmetric and asymmetric kinetic terms (KTs) of the hamiltonian,
impose patterns of three order parameters (or triplets of order
parameters) that are unavoidable. Whenever two of the order
parameters coexist the third one appears as well. Therefore, we
must necessarily consider all three order parameters
simultaneously.
\par We have considered a tetragonal tight binding system and we have
studied 16 cases of phase coexistence involving various types of
ferromagnetism, density waves
\cite{Gruner,{Berlinsky},{Schulz},{Thalmeier},{Nayak1}} and
superconductivity \cite{Sigrist1}. In the case of a tetragonal
lattice, the possible OPs that can be observed along with the
kinetic terms are 63, generating an $SU(8)$ Lie algebra
\cite{Solomon}. The OPs we have studied, were chosen among those
of an $SO(8)$ subalgebra \cite{Markiewicz,{W.M.Zhang}} that
describes only even parity order parameters and is relevant for
the nearly half-filled case. We have observed that the kinetic
terms impose phases that do not initially exist in the
Hamiltonian. These phases ought to have been already included in
the Hamiltonian from the very beginning in order to study the
system consistently. We have performed this procedure in all cases
that we have studied and we obtained in all cases the
self-consistence equations of all the order parameters involving
the induced order. We observed that the two initial order
parameters, the induced order and the mixing kinetic term satisfy
a system of self-consistence equations that entangles their
dynamics. They constitute closed sets of order parameters that
need to be treated on the same footing.
\par Through detailed examination of a number of systems with many
phases emphasizing on the role of particle-hole symmetric and
asymmetric kinetic terms, we managed to extract a simple empirical
rule which helped us to predict patterns of OPs that coexist when
the kinetic terms are properly taken into account.
\textit{According to this rule the matrix product of the matrix
representations of the two initially coexisting order parameters
and the mixing kinetic term yields the matrix representation of
the induced phase. Equivalently, the matrix product of the three
involved order parameters yields the mixing KT matrix.} In all
cases that were selected according to this rule the predicted
phase coexistence was confirmed.
\par We have to remark that
although the above rule is quite expected if we want a non zero
mean value for the induced phase, there is no way to be certain
that if this rule stands for a specific set of order parameters
then we must obtain the coexistence described above unless we do
the calculations. Indeed, only by considering on a suitable spinor
formalism the relevant triplets of order parameters within a BCS
like mean field approach and extracting self-consistent gap
equations we were able to identify definitely that these states
appear altogether as an unavoidable pattern.

\section{\label{section:2}Results}

We consider all even parity OPs that are possible in a tetragonal
system and may be relevant for discussing a number of heavy
fermion materials as well as high-T$_c$ cuprates. To describe in a
unified way the coexistence of various OPs we need to introduce an
eight component spinor formalism.  We introduce the following
spinor \bea\fl\Psi^{\dag}_{{\bi{k}}}=
\left(\begin{array}{cccccccc}
\alpha^{\dagger}_{{\bi{k}}\uparrow}&\alpha^{\dagger}_{{\bi{k}}\downarrow}&
\alpha_{{\bi{k+Q}}\uparrow}^{\dag}&\alpha_{{\bi{k+Q}}\downarrow}^{\dag}&
\alpha^{\phd}_{{\bi{-k}}\uparrow}&\alpha^{\phd}_{{\bi{-k}}\downarrow}&
\alpha_{{\bi{-k-Q}}\uparrow}^{\phd} &
\alpha_{{\bi{-k-Q}}\downarrow}^{\phd}\\\end{array}\right)\end{eqnarray}

\noi where $\alpha_{\bi{k}s}^{\phd}/\alpha_{\bi{k}s}^{\dag}$ are
the destruction/creation operators of an electron of momentum
$\bi{k}$ in the Reduced Brillouin zone and spin projection
$s=\uparrow,\downarrow$. This enlargement of the spinor space
allows the simultaneous description of ferromagnetism, zone center
(zero Cooper-pair momentum) and staggered (finite Cooper-pair
momentum) superconductivity, charge and spin density waves. The
density waves and the staggered superconductivity are
characterized by the wave-vector $\bi{Q}=(\pi,\pi)$ which is the
best nesting vector close to half-filling. To work in this eight
dimensional spinor space we consider a base formed by the
Kronecker products of the unit matrix and three of the usual Pauli
matrices $\tau_i^{\phd},\rho_j^{\phd},\sigma_k^{\phd}$ where
$i,j,k=1,2,3$.

\par
The possible order parameters arising from the preceding spinor
theory are $4\times4\times4=64$. If we demand that our Hamiltonian
is traceless we are left with 63 order parameters (including K.T.)
that constitute the generators of an $SU(8)$ spectrum generating
algebra \cite{Solomon}. In the case of tetragonal systems close to
half-filling, equivalence of the Brillouin zone points
$(\pi,0)\equiv((-\pi,0))$ and $(0,\pi)\equiv(0,-\pi)$ imposes that
the order parameters have even parity. The OPs satisfying this
constraint are 28 (including K.T.) and form an $SO(8)$ spectrum
generating algebra \cite{Markiewicz,W.M.Zhang}. The OPs that we
have considered in this study were chosen among those 28
identified in \Tref{table:SO(8)} with their symbols adopted here.

\begin{table}[!hbp]
\caption{\label{table:SO(8)}The 28 OPs that form an $SO(8)$
spectrum generating algebra and would be accessible in a
tetragonal system close to half filling. In the next sections we
demonstrate that particle-hole asymmetric and symmetric kinetic
terms impose various patterns of triplets of the following OPs.}

\begin{indented}
\item[]\begin{tabular}{@{}ll} \br
\textbf{\textit{Order Parameter}} &\textbf{\textit{Type}}\\
\mr
$\gamma$&{nearest neighbours hopping term}\\
$\delta$&{next nearest neighbours hopping term}\\
$F_{x,y,z}^{\phd}$&{ferromagnet along x,y,z-axis}\\
$A_{x,y,z}^{\phd}$&{d-wave ferromagnet along x,y,z-axis}\\
$W$&{charge density wave}\\
$J_c^{\phd}$&{orbital anti-ferromagnet}\\
$M_{x,y,z}^{\phd}$&{spin density wave along x,y,z-axis}\\
$J^s_{x,y,z}$&{spin nematic along x,y,z-axis}\\
$\Delta_s^{\phd}$&s-wave SC ($\bi{q}=0$)\\
$\Delta_d^{\phd}$&d-wave SC ($\bi{q}=0$)\\
$\eta^{\phd}$&s-wave SC ($\bi{q}=\bi{Q}$)\\
$\Pi_{x,y,z}^{\phd}$&{d-wave SC along x,y,z-axis} ($\bi{q}=\bi{Q}$)\\
\br
\end{tabular}
\end{indented}
\end{table}

We note that in \Tref{table:SO(8)} there are 16 OPs corresponding
to particle-hole condensates including the KTs and 12
superconducting states (including their complex conjugates).
Moreover, 8 of the 12 superconducting OPs represent staggered SC
in which the pairs have a finite total center-of-mass momentum
$(\bi{q}=\bi{Q})$ bearing similarities to the Fulde-Ferrel states
\cite{Fulde}. These quite exotic states are superconducting states
with modulated superfluid density and as we will show below, they
should play a crucial role in any antiferromagnetic SC state.
\par We report here 16 different
patterns of OPs that are imposed by the particle-hole symmetric
and asymmetric kinetic terms. They can be classified into 3
different types of OP mixing that according to their properties
can be merged into two general groups. In all these cases we
present the typical system of self-consistence equations that
provide the OPs and we identify the kinetic terms that are
responsible for the OPs mixing.

\subsection{\label{subsection:1}First type of OPs mixing}

In this first case, we consider that the Hamiltonian consists of
the two kinetic terms and three order parameters. These order
parameters have been chosen according to the empirical rule
mentioned in the introduction i.e. the matrix product of the three
order parameters yields the kinetic term that causes their mixing.
In \Tref{table:case1} we present the different combinations that
fall into this class.

\begin{table}[!hbp]
\caption{\label{table:case1}Triplets of order parameters that form
patterns imposed by the kinetic terms mentioned in the last
column. In all cases we have the same system of self-consistence
equations for the OPs, provided we replace the corresponding OPs
of the same column.\\}
\begin{indented}
\item[]\begin{tabular}{@{}llll}\br
\it{\bt{OP}} 1&\it{\bt{OP}} 2&\it{\bt{OP}} 3&\it{\bt{mixing KT}}\\
\mr
$M_z^{\phd}$&$\Delta_d^{\phd}$&$\Pi_z^{\phd}$&$\delta$\\
$W$&$\Delta_s^{\phd}$&$\eta$&$\delta$\\
$J^s_{y}$&$\Delta_s^{\phd}$&$\Pi_y^{\phd}$&$\delta$\\
$\Pi_z^{\phd}$&$\Delta_s^{\phd}$&$M_z^{\phd}$&$\gamma$\\
$\eta$&$\Delta_d^{\phd}$&$W$&$\gamma$\\
$\Pi_y^{\phd}$&$\Delta_d^{\phd}$&$J^s_{y}$&$\gamma$\\
\br
\end{tabular}
\end{indented}
\end{table}

We consider explicitly the first combination of the preceding
table in order to demonstrate the general equations governing the
phase coexistence and competition in the above patterns. In fact,
the self-consistence equations that result are the same for all
six patterns provided we replace the corresponding OPs that are in
the same column. The Hamiltonian corresponding to the first case
is given by the relation

\bea\fl
 H=\sum_{\bi{k}} \Psi_{{\bi{k}}}^\dagger \left( \gamma
\tau_3^{\phd}\rho_3^{\phd}+\delta\tau_3^{\phd}-M_z^{\phd}
\tau_3^{\phd}\rho_1^{\phd}\sigma_3^{\phd}-\Pi_z^{\phd}
\tau_2^{\phd}\rho_2^{\phd}\sigma_1^{\phd}+\Delta_d^{\phd}\tau_2^{\phd}\rho_3^{\phd}
\sigma_2^{\phd}\right)\Psi_{{\bi{k}}}\eea

\noi where we have suppressed the momentum index $\bi{k}$ and the
Kronecker product's symbol $\otimes$. The energy eigenvalues are

\begin{eqnarray}\fl
E_\pm=\sqrt{M_z^2+\gamma^2+\delta^2+\Pi_z^2+\Delta_d^2\pm2
\sqrt{(M_z^2+\gamma^2)\delta^2-2\delta M
_z^{\phd}\Pi_z^{\phd}\Delta_d^{\phd}+
(\Delta_d^2+\gamma^2)\Pi_z^2}}
\end{eqnarray}

\noi The self-consistence equations of the order parameters are
the following

\begin{eqnarray}
\fl M_z^{\phd}=\frac{1}{4}\sum_{\bi{k'}}\frac{V_{{\bi{k
}}{\bi{k'}}}^{M_z^{\phd}}}{E_+^2-E_-^2}
\left\{\ph\frac{M_z^{\phd}\left(E_+^2-E_-^2+4\delta^2\right)-
4\delta\Delta_d^{\phd}\Pi_z^{\phd}}{E_+}\tanh\left(\frac{E_+}{2T}\right)\right.\no\\
\left.+\frac{M_z^{\phd}\left(E_+^2-E_-^2-4\delta^2\right)+
4\delta\Delta_d^{\phd}\Pi_z^{\phd}}{E_-}\tanh\left(\frac{E_-}{2T}\right)\right\}\\
\fl\Delta_d^{\phd}=\frac{1}{4}\sum_{\bi{k'}}\frac{V_{{\bi{k
}}{\bi{k'}}}^{\Delta_d^{\phd}}}{E_+^2-E_-^2}\left\{\ph\frac{\Delta_d^{\phd}\left(E_+^2-E_-^2+4\Pi_z^2\right)-
4\delta M_z^{\phd}\Pi_z^{\phd}}{E_+}\tanh\left(\frac{E_+}{2T}\right)\right.\no\\
\left.+\frac{\Delta_d^{\phd}\left(E_+^2-E_-^2-4\Pi_z^2\right)+4\delta
M_z^{\phd}\Pi_z^{\phd}}{E_-}\tanh\left(\frac{E_-}{2T}\right)\right\}\\
\fl\Pi_z^{\phd}=\frac{1}{4}\sum_{\bi{k'}}\frac{V_{{\bi{k}}{\bi{k'}}}^{\Pi_z^{\phd}}}{E_+^2-E_-^2}
\left\{\ph\frac{\Pi_z^{\phd}\left(E_+^2-
E_-^2+4\gamma^2+4\Delta_d^2\right)-4\delta\Delta_d^{\phd}M_z^{\phd}}{E_+}\tanh\left(\frac{E_+}{2T}\right)\right.\no\\
\left.+\frac{\Pi_z^{\phd}\left(E_+^2-E_-^2-4\gamma^2-4\Delta_d^2\right)+
4\delta\Delta_d^{\phd}M_z^{\phd}}{E_-}\tanh\left(\frac{E_-}{2T}\right)\right\}\eea

\noi where the OPs and KTs depend on $\bi{k'}$. The mixing role of
the kinetic term is explicit already in the form of the equation.
In fact, the usual BCS equations for each one of the order
parameters are expected to have the general form

\bea
\Delta_{\bi{k}}^{\phd}=\sum_{\bi{k'}}f(E_{\bi{k'}}^{\phd},T)V_{\bi{k}\bi{k'}}^{\phd}\Delta_{\bi{k'}}^{\phd}
\eea

\noi where $f$ is a function of the energy dispersion $E$ and
temperature $T$. Each BCS equation supports solutions of zero and
non-zero order parameter, depending on the temperature. On the
other hand, in our case each OP self-consistence equation has the
general form

\bea
\Delta_{\bi{k}}^{\phd}=\sum_{\bi{k'}}V_{\bi{k,k'}}^{\phd}\left\{f(E_{\bi{k'}}^{\phd},T)\Delta_{\bi{k'}}^{\phd}
+g(E_{\bi{k'}}^{\phd},T)m_{\bi{k'}}^{\phd}A_{\bi{k'}}^{\phd}B_{\bi{k'}}^{\phd}\right\}\label{eq:self-consistence}\eea

\noi where $f,g$ are function of the energy dispersion and
temperature, $m_{\bi{k}}^{\phd}$ is the mixing kinetic term and
$A_{\bi{k}}^{\phd}$ and $B_{\bi{k}}^{\phd}$ are the other two OPs.
\textit{We observe that a solution of zero order parameter is not
possible unless the mixing term or one at least of the other order
parameters is also zero.} \bt{\textit{This suggests that in the
presence of the mixing kinetic term we cannot have two order
parameters without the third.}} \textit{The three order parameters
and the mixing term constitute a group that must be treated as an
independent subsystem on the same footing.} \par It is interesting
to obtain the self-consistence equations when the kinetic term
that does not contribute to the mixing is set to zero. In this
case, the eigenenergies obtain the form

\bea
E_{+}^{\phd}&=&\sqrt{(M_{z}^{\phd}+\delta)^2+(\Delta_d^{\phd}-\Pi_{z}^{\phd})^2}\\
E_{-}^{\phd}&=&\sqrt{(M_{z}^{\phd}-\delta)^2+(\Delta_d^{\phd}+\Pi_{z}^{\phd})^2}\eea

\noi while the first self-consistence equation becomes

\bea\fl M_z^{\phd}=\sum_{\bi{k'}}V_{{\bi{k
}}{\bi{k'}}}^{M_z^{\phd}}\left\{\frac{M_z^{\phd}+\delta}{\sqrt{(M_{z}^{\phd}+\delta)^2+(\Delta_d^{\phd}-\Pi_{z}^{\phd})^2}}
\tanh\left(\frac{E_+}{2T}\right)\right.\no\\
\left.+\frac{M_z^{\phd}-\delta}{\sqrt{(M_{z}^{\phd}-\delta)^2+(\Delta_d^{\phd}+\Pi_{z}^{\phd})^2}}
\tanh\left(\frac{E_-}{2T}\right)\right\}\eea

\subsection{\label{subsection:2}Second type of OPs mixing}

The second case involves a different coexistence pattern involving
once again three order parameters and a mixing kinetic term. The
following table contains the combinations belonging in this class:

\begin{table}[!hbp]
\caption{\label{table:case2}Same as in \Tref{table:case1}}
\begin{indented}
\item[]
\begin{tabular}{@{}llll}\br
\it{\bt{OP}} 1&\it{\bt{OP}} 2&\it{\bt{OP}} 3&\it{\bt{mixing KT}}\\
\mr
$F_z^{\phd}$&$\Pi_x^{\phd}$&$\Pi_y^{\phd}$&$\delta$\\
$A_z^{\phd}$&$\eta$&$\Pi_z^{\phd}$&$\delta$\\
$F_y^{\phd}$&$J^s_{x}$&$M_z^{\phd}$&$\gamma$\\
$A_z^{\phd}$&$W$&$M_z^{\phd}$&$\gamma$\\
\br
\end{tabular}
\end{indented}
\end{table}

\noi Once again we present the typical results of one of these
cases, specifically we consider the first. The Hamiltonian is

\begin{equation}
 H=\sum_{\bi{k}} \Psi_{\bi{k}}^\dagger \left( \gamma
\tau_3^{\phd}\rho_3^{\phd} +\delta \tau_3^{\phd}
-\Pi_y^{\phd}\tau_2^{\phd}\rho_2^{\phd} -\Pi_x^{\phd}
\tau_2^{\phd}\rho_2^{\phd}\sigma_3^{\phd} -F_z^{\phd}
\tau_3^{\phd} \sigma_3^{\phd}  \right)\Psi_{\bi{k}}
\end{equation}

\noi The corresponding quasi-particle poles are

\begin{eqnarray}
E_{\pm+}=\gamma\pm\sqrt{\left(F_z^{\phd}- \delta
\right)^2+\left(\Pi_x^{\phd}+\Pi_y^{\phd}\right)^2}\\
E_{\pm-}=\gamma\pm\sqrt{\left(F_z^{\phd}+\delta
\right)^2+\left(\Pi_x^{\phd}-\Pi_y^{\phd}\right)^2}
\end{eqnarray}

\noi Finally, we obtain the self-consistence equations

\begin{eqnarray}
\fl F_z^{\phd}=\frac{1}{8}\sum_{\bi{k'}}V_{{\bi{k
}}{\bi{k'}}}^{F_z^{\phd}}\left\{\frac{F_z^{\phd}-\delta}{\sqrt{\left(F_z^{\phd}-\delta\right)^2+
\left(\Pi_x^{\phd}+\Pi_y^{\phd}\right)^2}}\left[\tanh\left(\frac{E_{++}}{2T}\right)-\tanh\left(\frac{E_{-+}}{2T}\right)\right]\right.\no\\
\left.+\frac{F_z^{\phd}+\delta}{\sqrt{\left(F_z^{\phd}+\delta\right)^2+\left({\Pi_x^{\phd}}-{\Pi_y^{\phd}}\right)^2}}
\left[\tanh\left(\frac{E_{+-}}{2T}\right)-\tanh\left(\frac{E_{--}}{2T}\right)\right]\right\}\quad\label{eq:Fz}\\
\fl\Pi_x=\frac{1}{8}\sum_{\bi{k'}}V_{{\bi{k
}}{\bi{k'}}}^{\Pi_x^{\phd}}\left\{
\frac{{\Pi_x^{\phd}}+{\Pi_y^{\phd}}}{\sqrt{\left({\Pi_x^{\phd}}+{\Pi_y^{\phd}}\right)^2+
\left(F_z^{\phd}-\delta\right)^2}}\left[\tanh\left(\frac{E_{++}}{2T}\right)-\tanh\left(\frac{E_{-+}}{2T}\right)\right]\right.\no\\
\left.+\frac{{\Pi_x^{\phd}}-{\Pi_y^{\phd}}}{\sqrt{\left({\Pi_x^{\phd}}-{\Pi_y^{\phd}}\right)^2+
\left(F_z^{\phd}+\delta\right)^2}}\left[\tanh\left(\frac{E_{--}}{2T}\right)-\tanh\left(\frac{E_{+-}}{2T}\right)\right]\right\}\quad\label{eq:Px}\\
\fl\Pi_y^{\phd}=\frac{1}{8}\sum_{\bi{k'}}V_{\bi{kk'}}^{\Pi_y^{\phd}}\left\{
\frac{{\Pi_x^{\phd}}+{\Pi_y^{\phd}}}{\sqrt{\left({\Pi_x^{\phd}}
+\Pi_y^{\phd}\right)^2+ \left(\delta -F_z^{\phd}\right)^2}}
\left[\tanh\left(\frac{E_{++}}{2T}\right)-\tanh\left(\frac{E_{-+}}{2T}\right)\right]\right.\no\\
\left.+\frac{{\Pi_x^{\phd}}-{\Pi_y^{\phd}}}{\sqrt{\left(\Pi_x^{\phd}
-\Pi_y^{\phd}\right)^2+\left(F_z^{\phd}+\delta\right)^2}}
\left[\tanh\left(\frac{E_{+-}}{2T}\right)-\tanh\left(\frac{E_{--}}{2T}\right)\right]\right\}
\end{eqnarray}

\noi As one can observe, the above equations do not have the
general form (\ref{eq:self-consistence}) in which the mixing role
of the relevant kinetic term is explicit as in the previous case.
However, by performing a Taylor expansion with respect to the
order parameters up to quadratic order terms we can show that such
a relation does exist. These self-consistence equations imply once
again that if the mixing term is present we cannot have the two
order parameters without the third.
\par Indeed, let us consider \eref{eq:Fz} supposing that
$F_z^{\phd}$ is absent from the initial Hamiltonian. Then we set
$F_z^{\phd}=0$ in the right side of \eref{eq:Fz} and this equation
will now provide the {\it induced} part of $F_z^{\phd}$:

\bea \fl F_z^{\ph
induced}\sim\sum_{\bi{k'}}\left\{\ph\frac{-\delta}{\sqrt{\delta^2+
\left(\Pi_x^{\phd}+\Pi_y^{\phd}\right)^2}}\left[\tanh\left(\frac{E_{++}'}{2T}\right)-\tanh\left(\frac{E_{-+}'}{2T}\right)\right]\right.\no\\
\left.+\frac{\delta}{\sqrt{\delta^2+\left({\Pi_x^{\phd}}-{\Pi_y^{\phd}}\right)^2}}
\left[\tanh\left(\frac{E_{+-}'}{2T}\right)-\tanh\left(\frac{E_{--}'}{2T}\right)\right]\right\}\eea

\noi where we have introduced the new energy dispersions $E'$, by
setting $F_z^{\phd}=0$:

\begin{eqnarray}
E_{\pm+}'=\gamma\pm\sqrt{\delta^2+\left(\Pi_x^{\phd}+\Pi_y^{\phd}\right)^2}\\
E_{\pm-}'=\gamma\pm\sqrt{\delta^2+\left(\Pi_x^{\phd}-\Pi_y^{\phd}\right)^2}
\end{eqnarray}

A zero induced $F_z^{\phd}$ term is expected if one of the two
following conditions holds. On one hand, we may have $\delta=0$,
in which case we confirm that we will not have induced
$F_z^{\phd}$ if the mixing kinetic term vanishes. On the other
hand, we may have $E_{\pm+}=E_{\mp+}$ and $E_{\pm-}=E_{\mp-}$.
This last condition can be realized only when $\Pi_x^{\phd}=0$ or
$\Pi_{y}^{\phd}=0$. \textit{Consequently we conclude that if the
mixing term is present and two order parameters are non zero we
have an induced order $F_{z}^{\phd}$. i.e. the three phases
coexist}.
\par It is instructive to derive here as well, the
self-consistence equations when the irrelevant, to the mixing,
kinetic term vanishes.

\begin{eqnarray}
\tilde{E}_{\pm+}=E_{\pm+}^{\gamma=0}=\pm\sqrt{\left(F_z^{\phd}-
\delta
\right)^2+\left(\Pi_x^{\phd}+\Pi_y^{\phd}\right)^2}\\
\tilde{E}_{\pm-}=E_{\pm-}^{\gamma=0}=\pm\sqrt{\left(F_z^{\phd}+\delta
\right)^2+\left(\Pi_x^{\phd}-\Pi_y^{\phd}\right)^2}
\end{eqnarray}

\noi We observe that $\tilde{E}_{+\pm}=-\tilde{E}_{-\mp}$. This
equality simplifies the self-consistence equations. For example we
have

\bea\fl F_z^{\phd}=\frac{1}{4}\sum_{\bf{i'}}V_{{\bi{k
}}{\bi{k'}}}^{F_z^{\phd}}\left\{\frac{F_z^{\phd}-\delta}{\sqrt{\left(F_z^{\phd}-\delta\right)^2+
\left(\Pi_x^{\phd}+\Pi_y^{\phd}\right)^2}}\tanh\left(\frac{\tilde{E}_{++}}{2T}\right)\right.\no\\
\left.+\frac{F_z^{\phd}+\delta}{\sqrt{\left(F_z^{\phd}+\delta\right)^2+\left({\Pi_x^{\phd}}-{\Pi_y^{\phd}}\right)^2}}
\tanh\left(\frac{\tilde{E}_{+-}}{2T}\right)\right\}\eea

\noi Quite remarkably, we have encountered the same form of
self-consistence equation in \Sref{subsection:1} when the non
mixing kinetic term was set to zero. \textit{This common feature
reveals that these two cases share the same mixing ``mechanism'',
constituting specific examples of a more general coexistence
pattern.}
\par The next pattern that we shall discuss here is the coexistence
of two specific phases, s-wave and d-wave SC OPs in the presence
of the two kinetic terms, \textit{where the kinetic terms play
both the role of the mixing terms and the OPs at the same time.}
As far as the form of the equations that we derive, they belong to
the same general coexistence pattern like the one reported just
above. The Hamiltonian is

\bea H=\sum_{\bi{k}}\Psi_{\bi{k}}^\dagger \left( \gamma
\tau_3^{\phd} \rho_3^{\phd}+\delta\tau_3^{\phd}-\Delta_d^{\phd}
\tau_2^{\phd}\rho_3^{\phd}\sigma_2^{\phd}-\Delta_s^{\phd}
\tau_2^{\phd}\sigma_2^{\phd} \right)\Psi_{\bi{k}}\eea

\noi The poles of the Green's function are

\bea E_{\pm}^{\phd}=\sqrt{\left(\Delta_s^{\phd}\pm
\Delta_d^{\phd}\right)^2+\left(\gamma\pm \delta\right)^2}\eea

\noi It is evident that they have the form of the previous cases
\Sref{subsection:1} and \Sref{subsection:2} when we set the
irrelevant kinetic terms equal to zero. The self-consistence
equations obey the same rule

\begin{eqnarray}
\fl\Delta_d^{\phd}=\frac{1}{4}\sum_{\bi{k'}}V_{{\bi{k
}}{\bi{k'}}}^{\Delta_d^{\phd}}\left\{\frac{\Delta_d^{\phd}+\Delta_s^{\phd}}{\sqrt{\left(\Delta_s^{\phd}
+\Delta_d^{\phd}\right)^2+\left(\gamma+
\delta\right)^2}}\tanh\left(\frac{E_{+ }}{2T}\right)\right.\no\\
\left.+\frac{{\Delta_d^{\phd}}-\Delta_s^{\phd}}{\sqrt{\left(\Delta_s^{\phd}
-\Delta_d^{\phd}\right)^2+\left(\gamma-\delta\right)^2}}\tanh
\left(\frac{E_{-}}{2T}\right)\right\}\no\\
\fl\Delta_s^{\phd}=\frac{1}{4}\sum_{\bi{k'}}V_{{\bi{k
}}{\bi{k'}}}^{\Delta_s^{\phd}}\left\{\frac{\Delta_s^{\phd}+\Delta_d^{\phd}}{\sqrt{\left(\Delta_s^{\phd}+\Delta_d^{\phd}\right)^2
+\left(\gamma+\delta\right)^2}}\tanh\left(\frac{E_{+}}{2T}\right)\right.\no\\
\left.+\frac{\Delta_s^{\phd}-\Delta_d^{\phd}}{\sqrt{\left(\Delta_s^{\phd}-\Delta_d^{\phd}\right)^2+\left(\gamma-
\delta\right)^2}}\tanh\left(\frac{E_{-}}{2T}\right)\right\}\end{eqnarray}

\subsection{\label{subsection:3}Third type of OPs mixing}

The next case we consider has distinct properties from the
preceding encountered in \Sref{subsection:1} and
\Sref{subsection:2}. We have found that the following combinations
all have the same coexistence pattern.

\begin{table}[!hbp]
\caption{\label{table:case3}Distinct type of mixing compared to
the one related to \Sref{subsection:1} and \Sref{subsection:2}.
The following combinations obey the same system of
self-consistence equations.\\}
\begin{indented}
\item[]\begin{tabular}{@{}llll}\br
\it{\bt{OP}} 1&\it{\bt{OP}} 2&\it{\bt{OP}} 3&\it{\bt{mixing KT}}\\
\mr
$F_z^{\phd}$&$W$&$M_z^{\phd}$&$\delta$\\
$F_z^{\phd}$&$J_c^{\phd}$&$J^s_z$&$\delta$\\
$A_x^{\phd}$&$J^s_{y}$&$M_z^{\phd}$&$\delta$\\
$F_z^{\phd}$&$\eta$&$\Pi_z^{\phd}$&$\gamma$\\
\br
\end{tabular}
\end{indented}
\end{table}

\noi The example in this type of mixing is the first of
\Tref{table:case3}, which is described by the Hamiltonian

\begin{equation}
 H=\sum_{\bi{k}}\Psi_{\bi{k}}^\dag\left(\gamma
\tau_3^{\phd}\rho_3^{\phd}+\delta\tau_3^{\phd}-M_z^{\phd}
\tau_3\rho_1^{\phd}\sigma_3-W\tau_3^{\phd}\rho_1^{\phd}-F_z^{\phd}
\tau_3^{\phd}\sigma_3^{\phd} \right)\Psi_{\bi{k}}
\end{equation}

\noi with corresponding eigenenergies

\begin{eqnarray}
E_{\pm+}^{\phd}=(F_z^{\phd}\mp\delta)+\sqrt{\left(M_z^{\phd}\pm
W\right)^2+\gamma^2}\\
E_{\pm-}^{\phd}=(F_z^{\phd}\mp\delta)-\sqrt{\left(M_z^{\phd}\pm
W\right)^2+\gamma^2}\end{eqnarray}

\noi We observe that the structure of the poles are different from
the ones found in Sections \ref{subsection:1} and
\ref{subsection:2}. The self-consistence equations are given from
the following relations

\begin{eqnarray}
\fl F_z^{\phd}=\frac{1}{8}\sum_{\bi{k'}}V_{\bi{kk'}}^{F_z^{\phd}}
\left\{\tanh\left(\frac{E_{++}}{2T}\right)+\tanh\left(
\frac{E_{+-}}{2T}\right)
+\tanh\left(\frac{E_{-+}}{2T}\right)+\tanh\left(\frac{E_{--}}{2T}\right)\right\}\label{eq:FF}\no\\\\
\fl
M_z^{\phd}=\frac{1}{8}\sum_{\bi{k'}}V_{\bi{kk'}}^{M_z^{\phd}}\left\{\frac{M_z^{\phd}+W}{\sqrt{\left(M_z^{\phd}+
W\right)^2+\gamma^2}}
 \left[\tanh\left(\frac{E_{++}}{2T}\right)-\tanh\left(\frac{E_{+-}}{2T}\right)
 \right]\right.\no\\
 \left.+\frac{M_z^{\phd}-W}{\sqrt{\left(M_z^{\phd}-W\right)^2+\gamma^2}}
 \left[\tanh\left(\frac{E_{-+}}{2T}\right)-\tanh\left(\frac{E_{--}}{2T}\right)\right]\right\}\label{eq:M}\\
\fl
W=\frac{1}{8}\sum_{\bi{k'}}V_{\bi{kk'}}^{W}\left\{\frac{M_z^{\phd}+W}
{\sqrt{\left(M_z^{\phd}+W\right)^2+\gamma^2}}
 \left[\tanh\left(\frac{E_{++}}{2T}\right)-\tanh\left(\frac{E_{+-}}{2T}\right)\right]\right.\no\\
 \left.+\frac{M_z^{\phd}-W}{\sqrt{\left(M_z^{\phd}-W\right)^2+\gamma^2}}
 \left[\tanh\left(\frac{E_{--}}{2T}\right)-\tanh\left(\frac{E_{-+}}{2T}\right)\right]\right\}\label{eq:W}
 \eea

\noi We have to remark that Equations \eref{eq:M} and \eref{eq:W}
have similarities with the results of the previous sections.
Though, \Eref{eq:FF} is totally different. Close observation of
the equation and the eigenenergies, shows that great
simplification occurs when $\gamma=0$. In this case we have

\begin{eqnarray}
E_{\pm+}^{\phd}=(F_z^{\phd}\mp\delta)+(M_z^{\phd}\pm W)\\
E_{\pm-}^{\phd}=(F_z^{\phd}\mp\delta)-(M_z^{\phd}\pm
W)\end{eqnarray}

\noi and as far as the self-consistence equations are concerned

\begin{eqnarray}
 \fl F_{z\phd}^{\phd}=\frac{1}{8}\sum_{\bi{k'}}V_{\bi{kk'}}^{F_{z\ph}^{\phd}}
\left\{\tanh\left(\frac{E_{++}}{2T}\right)+\tanh\left(
\frac{E_{+-}}{2T}\right)\right.\left.+\tanh\left(\frac{E_{-+}}{2T}\right)+\tanh\left(\frac{E_{--}}{2T}\right)\right\}\no\\\\
\fl
M_z^{\phd}=\frac{1}{8}\sum_{\bi{k'}}V_{\bi{kk'}}^{M_z^{\phd}}\left\{
 \left.\tanh\left(\frac{E_{++}}{2T}\right)-\tanh\left(\frac{E_{+-}}{2T}\right)
 +\tanh\left(\frac{E_{-+}}{2T}\right)-\tanh\left(\frac{E_{--}}{2T}\right)\right.\right\}\no\\\\
\fl
W_{\phd}^{\phd}=\frac{1}{8}\sum_{\bi{k'}}V_{\bi{kk'}}^{W\ph}\left\{
 \left.\tanh\left(\frac{E_{++}}{2T}\right)-\tanh\left(\frac{E_{+-}}{2T}\right)-
 \tanh\left(\frac{E_{-+}}{2T}\right)+\tanh\left(\frac{E_{--}}{2T}\right)\right.\right\}\no\\
\eea

According to what he have been taught from the previous sections,
the symmetry the above equations present, implies that we have
reached to a triplet of order parameters that necessarily coexist
in the presence of the corresponding mixing kinetic term. This can
be shown as follows. Any of these order parameters can be zero
only if the spectrum is particle-hole symmetric. This occurs
\textit{only} when two out of the four terms are zero.
\textit{Consequently if three of these terms are non zero the
fourth will be non zero, too.}
\par The final case we present, has two order parameters and the two
kinetic terms. As we shall see it belongs to the same coexistence
pattern of the above cases. The Hamiltonian is

\bea
 H=\sum_{\bi{k}}\Psi_{\bi{k}}^\dagger \left( \gamma
\tau_3^{\phd}\rho_3^{\phd}+\delta\tau_3^{\phd}+A _x^{\phd}
\tau_3^{\phd}\rho_3^{\phd}\sigma_1^{\phd}+ F_x^{\phd}
\tau_3^{\phd}\sigma_1^{\phd}\right)\Psi_{\bi{k}}\eea

\noi The eigenenergies are equal to

\bea E_{\pm+}=(A _x^{\phd}\pm {F_x^{\phd}})+(\delta\pm\gamma)\\
E_{\pm-}=(A_x^{\phd}\pm F_x^{\phd})-(\delta\pm\gamma) \eea

\noi The corresponding self-consistence equations are

\begin{eqnarray}
\fl A_x^{\phd}=\frac{1}{8}\sum_{\bi{k'}}V_{{\bi{k
}}{\bi{k'}}}^{A_x^{\phd}}\left\{\tanh
\left(\frac{E_{++}}{2T}\right)+\tanh\left(\frac{E_{+-}}{2T}\right)
+\tanh\left(\frac{E_{-+}}{2T}\right)+\tanh\left(\frac{E_{--}}{2T}\right)\right\}\no\\\\
\fl
F_x^{\phd}=\frac{1}{8}\sum_{\bi{k'}}V_{{\bi{k}}{\bi{k'}}}^{F_x^{\phd}}
\left\{\tanh\left(\frac{E_{++}}{2T}\right)-\tanh\left(\frac{E_{+-}}{2T}\right)
+\tanh\left(\frac{E_{-+}}{2T}\right)-\tanh\left(\frac{E_{--}}{2T}\right)\right\}\no\\
\end{eqnarray}

\section{Discussion}

Having studied 16 cases of coexisting OPs in the presence of the
particle-hole symmetric and asymmetric KTs, we have found 3
different types of coexistence patterns. The 11 cases studied in
Sections \ref{subsection:1} and \ref{subsection:2} seem to have
the same type of coexistence ``mechanism'' belonging to a more
general \textit{Coexistence Scheme}. On the other hand the 5 cases
that we presented in \Sref{subsection:3} originate by a distinct
coexistence ``mechanism''. \bt{Consequently, we conclude that the
16 cases we studied merge into two general \textit{Coexistence
Schemes}} as in \Tref{table:General}.
\par Moreover, we have found that
special coexistence patterns can be observed even when he have
only 2 OPs. In this case the kinetic terms play a dual role. They
behave as the mixing KTs and the members of triplets of coexisting
OPs. In that case, with both kinetic terms present we cannot
observe one of the two OPs without the second one. For example, in
the presence of both kinetic terms, d-wave SC coexists with s-wave
SC ! This shows that the generation of these patterns is not a
special property owned by the KTs.\textbf{ Consequently, these
\textit{Coexistence Schemes} originate due to more general
relations that are satisfied by quartets of the $\bi{SU(8)}$
generators.} We expect that other terms, apart from the KTs could
play the role of the mixing terms and produce different type of
coexistence patterns. We will present elsewhere a complete account
of all patterns of coexisting states that correspond to the above
mentioned quartets.
\par Finally, apart from the general conclusions that we inferred
about phase coexistence, we obtained valuable results concerning
specific coexisting triplets of OPs that may correspond to the
physical situation in numerous correlated systems of interest.
Particularly, we have observed that:

\begin{itemize}
 \item Density waves, zone-center superconductivity $(\bi{q}=\bi{0})$ and
staggered super\-conductivity $(\bi{q}=\bi{Q})$ constitute a
triplet of OPs that necessarily coexist in the presence of the
KTs. Such an observation is of general relevance for all
antiferromagnetic superconductors, a category of materials that
includes organics, heavy fermions, high-$T_c$ cuprates etc.

\item Ferromagnetism, charge density waves and spin density waves
constitute another triplet of OPs that necessarily coexist in the
presence of the asymmetric KT. This observation has already been
reported before and shown to be related with the colossal
magnetoresistance phenomenon \cite{CMRprl}.

\item s-wave and d-wave superconducting OPs always coexist in the
combined presence of the symmetric and asymmetric KTs. Needless to
note that high-$T_c$ cuprates as well as numerous heavy fermion
systems are believed to be d-wave SC. Our observations imply that a
pure d-wave SC state is an oversimplification.

\item s-wave and d-wave ferromagnetic OPs always coexist in the
combined presence of the symmetric and asymmetric KTs. The
implications of this observation need to be investigated.
\end{itemize}

\ack We thank the European Union for financial support through the
STRP NMP4-CT-2005-517039 CoMePhS project. P.K. acknowledges
financial support by the Greek Scholarships State Foundation.

\begin{table}[t]
\caption{\label{table:General}\it{The two distinct \bt{Coexistence
Schemes}}\\}
\begin{indented}
\item[]\begin{tabular}{@{}lllll}\br
\it{\bt{Scheme}}&\it{\bt{OP}} 1&\it{\bt{OP}} 2&\it{\bt{OP}} 3&\it{\bt{mixing KT}}\\
\mr
\it{\bt{I}}&$M_z^{\phd}$&$\Delta_d^{\phd}$&$\Pi_z^{\phd}$&$\delta$\\
\it{\bt{I}}&$W$&$\Delta_s^{\phd}$&$\eta$&$\delta$\\
\it{\bt{I}}&$J^s_{y}$&$\Delta_s^{\phd}$&$\Pi_y^{\phd}$&$\delta$\\
\it{\bt{I}}&$\Pi_z^{\phd}$&$\Delta_s^{\phd}$&$M_z^{\phd}$&$\gamma$\\
\it{\bt{I}}&$\eta$&$\Delta_d^{\phd}$&$W$&$\gamma$\\
\it{\bt{I}}&$\Pi_y^{\phd}$&$\Delta_d^{\phd}$&$J^s_{y}$&$\gamma$\\
\it{\bt{I}}&$F_z^{\phd}$&$\Pi_x^{\phd}$&$\Pi_y^{\phd}$&$\delta$\\
\it{\bt{I}}&$A_z^{\phd}$&$\eta$&$\Pi_z^{\phd}$&$\delta$\\
\it{\bt{I}}&$F_y^{\phd}$&$J^s_{x}$&$M_z^{\phd}$&$\gamma$\\
\it{\bt{I}}&$A_z^{\phd}$&$W$&$M_z^{\phd}$&$\gamma$\\
\it{\bt{I}}&$\Delta_s^{\phd}$&$\Delta_d^{\phd}$&$\delta$&$\gamma$\\
\it{\bt{II}}&$F_z^{\phd}$&$W$&$M_z^{\phd}$&$\delta$\\
\it{\bt{II}}&$F_z^{\phd}$&$J_c^{\phd}$&$J^s_z$&$\delta$\\
\it{\bt{II}}&$A_x^{\phd}$&$J^s_{y}$&$M_z^{\phd}$&$\delta$\\
\it{\bt{II}}&$F_z^{\phd}$&$\eta$&$\Pi_z^{\phd}$&$\gamma$\\
\it{\bt{II}}&$F_x^{\phd}$&$A_x^{\phd}$&$\delta$&$\gamma$\\
\br
\end{tabular}
\end{indented}
\end{table}

\end{document}